\documentclass{article}
\usepackage[a4paper, left=1.5cm, right=1.5cm, top=2.5cm, bottom=2.5cm]{geometry}
\usepackage{graphicx} 
\usepackage{amsmath}
\usepackage{amssymb}
\usepackage{xcolor}
\usepackage{authblk} 
\usepackage{cite}

\title{Novel Solar System Probes for Primordial Black Holes}

\author[1]{Oem Trivedi\thanks{oem.trivedi@vanderbilt.edu}}
\author[2]{Abraham Loeb\thanks{aloeb@cfa.harvard.edu}}

\affil[1]{Department of Physics and Astronomy, Vanderbilt University, Nashville, TN 37235, USA}
\affil[2]{Astronomy Department, Harvard University, 60 Garden St., Cambridge, MA 02138, USA}

\date{\today}

\begin{document}

\maketitle

\begin{abstract}
Primordial Black Holes (PBHs) represent one of the more interesting ways to address dark matter, at the interface of both cosmology and quantum gravity. It is no surprise then that testing PBHs is a venue of active interest, with several cosmological and astrophysical probes constraining different mass ranges. In this work, we propose novel Solar System scale searches for PBHs, motivated by the unique precision and coverage of local observables. We show that asteroid to dwarf planet mass PBHs can induce measurable dipolar timing signatures in pulsar timing arrays, while planetary mass PBHs can generate detectable ADAF accretion flares through interactions with Kuiper Belt bodies. Together, these complementary approaches open a new observational frontier for probing PBHs across mass ranges that remain unconstrained by conventional cosmological methods.
\end{abstract}

\section{Introduction}
The true nature of dark matter is perhaps one of the greatest mysteries in modern physics. Despite the overwhelming astrophysical and cosmological evidence for its existence, ranging from galactic rotation curves and gravitational lensing to the cosmic microwave background anisotropies, the fundamental nature of dark matter remains unknown. Over the past decades, numerous candidates have been proposed, including weakly interacting massive particles (WIMPs), axions, sterile neutrinos and other exotic possibilities \cite{dm11rubin1970rotation,dm1Cirelli:2024ssz,dm2Arbey:2021gdg,dm3Balazs:2024uyj,dm4Eberhardt:2025caq,dm5Bozorgnia:2024pwk,dm6Misiaszek:2023sxe,dm7OHare:2024nmr,dm8Adhikari:2022sbh,dm9Miller:2025yyx}. Still, all experimental and observational searches have not resulted in any conclusive consensus which promts the community to explore more unconventional alternatives. This has made this problem occupy a central position at the intersection of particle physics and cosmology.
\\
\\
Among the most intriguing and long standing candidates for dark matter are primordial black holes (PBHs) which are hypothetical black holes formed in the early Universe due to the collapse of large density perturbations. PBHs require no new particle physics beyond gravity which makes them a minimal and elegant possibility \cite{pbh1pbhzel1966hypothesis,pbh2hawking1971gravitationally,pbh3carr1974black,pbh4carr1975primordial,pbh5chapline1975cosmological,pbh6hawking1975particle,pbh7hawking1974black}. Depending on their formation mechanism and subsequent evolution, PBHs can span an enormous mass range right from subatomic scales to stellar or even supermassive black holes. Their phenomenology connects early Universe physics to present day astrophysical observables, with implications for inflation, baryogenesis, and structure formation \cite{pbh8khlopov1980primordial,pbh9polnarev1985cosmology,pbh10khlopov2010primordial,pbh11carr2016primordial,dm10Trivedi:2025vry,pbh12carr2020primordial,pbh13carr2024observational}. In recent years, renewed interest in PBHs has been sparked by the detections of binary black hole mergers by gravitational wave observatories, as well as by the growing recognition that PBHs could naturally account for some or all of the dark matter abundance if their masses lie within specific allowed mass windows \cite{pbh14green2024primordial}.
\\
\\
Given their wide mass spectrum and potential cosmological relevance, there has been an extensive and ongoing effort to detect or constrain PBHs using a variety of observational and experimental strategies. These include microlensing surveys, cosmic microwave background spectral distortions, dynamical heating of stellar systems, and the search for Hawking radiation signatures. However, each of these methods faces its own limitations as for example, microlensing approaches depend sensitively on the assumed velocity and spatial distributions of compact objects, while Hawking radiation searches lose sensitivity for PBHs above certain mass thresholds. In this work, we propose novel solar system level tests for PBHs, discussing separately planetary and non-planetary mass PBHs. In Section 2 we give an overview of various PBH probes so far and why solar system tests are a viable route to look out for. In Section 3 we show how one can use Pulsar Timing Array (PTA) observations for PBHs while in Section 4 we discuss how ADAF accretion flares can be useful for PBH constraints, and then we finally conclude our work in Section 5.
\\
\\
\section{Need for Solar System Probes}
Primordial black holes can be probed across a wide range of mass scales through a diverse set of astrophysical and cosmological observations. Each method relies on a particular physical effect arising from the presence of compact objects, such as gravitational lensing, accretion emission, dynamical heating or gravitational wave signatures \cite{test1Cai:2022kbp,test2Niikura:2019kqi,test3Tashiro:2008sf,test4Graham:2023unf,test5Auffinger:2022khh,test6Klipfel:2025bvh}. One of the earliest and most direct ways to constrain PBHs is through microlensing surveys, where the temporary magnification of background stars reveals the presence of compact lenses. The characteristic Einstein radius and event duration are given by
\begin{equation}
R_{\rm E}=\sqrt{\frac{4GM}{c^2}\frac{D_{\rm L}(D_{\rm S}-D_{\rm L})}{D_{\rm S}}}, \qquad 
t_{\rm E}=\frac{R_{\rm E}}{v_{\rm rel}}
\end{equation}
where $D_{\rm L}$ and $D_{\rm S}$ are the lens and source distances respectively and $v_{\rm rel}$ is the relative transverse velocity. Surveys such as EROS, OGLE, and Subaru–HSC have placed strong constraints in the mass range $10^{22}$–$10^{34}$ g, which leads to excluding PBHs as the dominant dark matter component over much of this interval \cite{test7Niikura:2017zjd,test8wyrzykowski2011ogle,test9EROS:2002flm}. But one should note here that microlensing methods lose sensitivity for both very light PBHs(where the lensing time is too short) and very heavy ones (where the event rate becomes exceedingly small), which ends up leaving significant open windows.
\\
\\
At smaller PBH masses another classic probe is that of Hawking radiation, as one can write that nonrotating PBH of mass $M$ emits particles thermally with a temperature
\begin{equation}
T_{\rm H}=\frac{\hbar c^3}{8\pi G M k_{\rm B}}
\end{equation}
and a corresponding luminosity is given by
\begin{equation}
L_{\rm H}\simeq \frac{\hbar c^6}{15360\pi G^2 M^2}
\end{equation}
For PBHs with $M\lesssim 10^{15}\ {\rm g}$, the emission is sufficiently intense to produce observable $\gamma$-ray fluxes which allwos for limits from the extragalactic $\gamma$-ray background and from nearby evaporating sources \cite{haw1Carr:2009jm,haw2Boudaud:2018hqb,haw3Laha:2019ssq}. Yet, above this mass threshold, the Hawking temperature becomes too low for any significant photon flux, effectively closing this detection channel. Furthermore, uncertainties in the cosmic PBH spatial distribution and initial mass function complicate direct inferences from the diffuse $\gamma$-ray signal.
\\
\\
At intermediate and high masses, accretion based and dynamical probes have been developed as well as PBHs embedded in baryonic environments can accrete gas which can, in principle, leave imprints on the cosmic microwave background as energy injection modifies recombination and reionization histories. Constraints from CMB anisotropies currently limit PBHs heavier than $10^{33}\ {\rm g}$ from constituting more than a small fraction of dark matter \cite{acc1Agius:2024ecw,acc2Serpico:2020ehh,acc3Piga:2022ysp}. Meanwhile, dynamical arguments based on the disruption of star clusters, wide binaries, and dwarf galaxies further restrict PBHs in the stellar mass regime.
\\
\\
Despite the breadth of these methods, significant portions of PBH parameter space remain weakly constrained as we see that each of the above probes depend on astrophysical environments and model dependent assumptions about accretion, clustering, and formation spectra. Furthermore, most observational techniques operate over cosmological or extragalactic baselines in which case the local, time-resolved effects of individual PBHs are averaged out. This leaves open the intriguing possibility that PBHs could still exist in abundance locally, which could be either be as isolated objects passing through the Solar neighborhood or even as gravitationally captured remnants within the Solar System. The characteristic number density of PBHs making up a fraction $f_{\rm PBH}$ of the local dark matter is given as
\begin{equation}
n_{\rm PBH}\simeq f_{\rm PBH}\frac{\rho_{\rm DM}}{M}
\end{equation}
which if we adopt the local masss density of  $\rho_{\rm DM}\sim 5\times10^{-25}\ {\rm g\,cm^{-3}}$ and $M\sim10^{20}$–$10^{30}\ {\rm g}$, implies inter-PBH separations of order au to $\mathcal{O}(10^3)$au, which thus makes nearby encounters potentially detectable.
\\
\\
Given this context, Solar System scale probes emerge as a compelling complementary approach as local gravitational effects from passing PBHs can, in principle, produce subtle yet measurable perturbations in well monitored astrophysical systems. Examples of this regard can be those of the barycenter motion of the Sun or the trajectories of planetary and small body populations. One can also see this in the sense that the characteristic acceleration given on target mass with separation $r$ from a PBH of mass $M_{\rm PBH}$ is
\begin{equation}
a_{\rm PBH}\sim \frac{G M_{\rm PBH}}{r^2}
\end{equation}
which for $M_{\rm PBH}\sim10^{25}\ {\rm g}$ and $r\sim 10^3$ au, gives us accelerations comparable to those measurable through precise timing and astrometric methods. In addition to this we see that the expected velocity perturbation over an observation time $T$ scales as $\Delta v\simeq a_{\rm PBH} T$, which allows us quite longer baselines for measurements to accumulate sufficient signal even for weak accelerations. With all this in mind, we will now discuss how we can use pulsar timing array data and accretion flares to probe PBHs in the Solar System.
\\
\\
\section{PTA Data and Asteroid-Mass PBHs}
Pulsar timing arrays are among the most sensitive gravitational observatories available today. By monitoring the times of arrival of radio pulses from millisecond pulsars distributed across the sky, PTAs can detect correlated variations induced by perturbations in the space-time metric. A passing gravitational wave or any other coherent gravitational perturbation that affects the Solar System barycenter, produces timing residuals with a characteristic angular correlation pattern across pulsars. Current PTA collaborations such as NANOGrav, EPTA, and PPTA have reached timing precision levels of tens of nanoseconds and have reported the detection of a stochastic background of nanohertz gravitational waves\cite{pta1NANOGrav:2023gor,pta2NANOGrav:2023hde,pta3Xu:2023wog,pta4Antoniadis:2022pcn,pta5Zic:2023gta,pta6Reardon:2023gzh,pta7EPTA:2023sfo}. With steadily improving sensitivity and extended time baselines, PTAs are now capable of constraining not only distant astrophysical sources but also low-frequency accelerations or perturbations affecting the Solar System itself and this is crucial as it opens the possibility of probing the gravitational effects of compact objects such as PBHs passing through or even just near the Solar neighborhood.
\\
\\
A PBH passing close to the Solar System imparts a small but finite gravitational impulse to the Solar System barycenter. This effect is analogous to a "velocity kick" and can be viewed as a sudden change in the Solar velocity relative to the pulsar reference frame. The induced change in velocity for a single encounter is \cite{Loeb:2024ekw}
\begin{equation}
\Delta v_{\odot} = \frac{2 G m_{\rm PBH}}{b v_{\rm PBH}}
\label{eq:dv_single}
\end{equation}
where $m_{\rm PBH}$ is the PBH mass, $v_{\rm PBH}$ its relative speed, and $b$ the impact parameter of the encounter and we can then take representative parameters for an asteroid-mass PBH,
\begin{equation}
m_{\rm PBH} = 10^{23}~{\rm g}, \quad v_{\rm PBH} = 2\times10^{7}~{\rm cm~s^{-1}}, \quad b = 1~{\rm au} = 1.5\times10^{13}~{\rm cm}
\end{equation}
With this, one can obtain
\begin{equation}
\Delta v_{\odot} \simeq 4.5\times10^{-5}~{\rm cm~s^{-1}}
\end{equation}
This velocity kick exceeds the typical stochastic velocity diffusion caused by asteroid interactions by roughly an order of magnitude and the associated fractional change in velocity with respect to the speed of light gives a characteristic strain-like amplitude,
\begin{equation}
h_{\rm PBH} \simeq \frac{\Delta v_{\odot}}{c} \simeq 1.5 \times 10^{-15}
\end{equation}
This corresponds to a correlated dipolar step in pulsar timing residuals that would manifest as a "memory-like" signal in the Solar System barycentric frame and is very close to the PTA sensitivity floor. The characteristic interaction timescale of the event, which is given by the flyby duration, is
\begin{equation}
\tau \sim \frac{b}{v_{\rm PBH}} \simeq 7.5\times10^{5}~{\rm s} \approx 8.7~{\rm days}
\end{equation}
Note that we have taken $b=1 $ au as an illustrative “nearby but non-catastrophic” pass to set a timescale for a resolvable kick. In practice we could have picked any b  with smaller b making kicks stronger but far rarer but in a cumulative random walk analysis
the b cancels out anyways. 
\begin{equation}
f \sim \frac{1}{\tau} \sim 1.3\times10^{-6}~{\rm Hz}
\end{equation}
This is slighty above the typical nanohertz PTA band but would still produce a permanent post-event offset visible in long term timing data.
\\
\\
When multiple PBHs traverse the Solar System vicinity over cosmological timescales, their cumulative effect can be modeled as a random walk in barycentric velocity space. If the mean rate of encounters within an impact parameter $b$ is $\Gamma$, the number of encounters in a total observing time $T$ is $N(T)\simeq \Gamma T$, then the rms barycentric velocity dispersion builds up as
\begin{equation} \label{sqrn}
\langle |\mathbf{v}_{\odot}|^{2}\rangle^{1/2} \simeq \sqrt{N(T)}\frac{2 G m_{\rm PBH}}{b v_{\rm PBH}} = \sqrt{\Gamma T}\frac{2 G m_{\rm PBH}}{b v_{\rm PBH}}
\end{equation}
This expression describes the diffusive, Brownian-like motion of the Solar System barycenter induced by repeated PBH encounters. The cumulative dipolar effect on PTA residuals would come through as a correlated red-noise component with a dipolar angular pattern which is quite distinct from isotropic gravitational wave backgrounds.
\\
\\
It is important to note that the $\sqrt{N}$ scaling used in \eqref{sqrn} refers to the 
temporal accumulation of discrete PBH encounters over an observing time $T$, 
where $N\simeq \Gamma T$ and each encounter contributes an independent kick 
$\Delta v \sim 2Gm_{\rm PBH}/(b v_{\rm PBH})$. As in any random walk process the rms 
barycentric drift therefore grows as
\begin{equation}
\langle |\mathbf{v}_{\odot}|^{2}\rangle^{1/2} \propto \sqrt{N}\,\Delta v
\end{equation}
which is the physically correct behavior for a sequence of uncorrelated impulses as $1/N$ or $1/\sqrt{N}$ temporal scaling would incorrectly diverge as $N\to 0$ which would be implying infinite kicks in the limit of no encounters, which is unphysical. By contrast, a $1/\sqrt{N}$ dependence appears only in the spatial continuum limit where one would fix the total DM density $\rho$ and increases the number of PBHs by taking $m_{\rm PBH}\to 0$, so that $n=\rho/m_{\rm PBH}\to\infty$. In this case the  granular gravitational noise would scale as 
\begin{equation}
\langle |\mathbf{v}_{\odot}|^{2}\rangle^{1/2} \propto m_{\rm PBH}^{1/2}
\propto \frac{1}{\sqrt{N_{\rm spatial}}}
\end{equation}
showing that a highly populated PBH field becomes smooth and fluid like. This limit,  
however, concerns spatial granularity and does not modify the $\sqrt{N}$ time domain  
diffusive behavior relevant for our PTA calculation here. 
\\
\\
The associated strain-like amplitude corresponding to this Brownian background is then given as
\begin{equation}
h_{c,{\rm PBH}}(T) \simeq \frac{\langle |\mathbf{v}_{\odot}|^{2}\rangle^{1/2}}{c} = \sqrt{\Gamma T}\frac{2 G m_{\rm PBH}}{b v_{\rm PBH} c}
\end{equation}
A null detection of such a correlated dipolar signature in PTA data provides an upper
limit on the PBH encounter rate $\Gamma$, and hence on the PBH number density $n_{\rm PBH}$,
through the geometric relation
\begin{equation}
\Gamma \simeq n_{\rm PBH}\,\pi b^{2} v_{\rm PBH}
\end{equation}
Inverting this expression yields
\begin{equation}
n_{\rm PBH} \lesssim 
\frac{1}{\pi b^{2} v_{\rm PBH} T}
\left(
\frac{h_{c,{\rm lim}}\, c\, b\, v_{\rm PBH}}{2G m_{\rm PBH}}
\right)^{2}
\label{eq:n_bound_new}
\end{equation}
where we adopt a conservative PTA dipolar sensitivity $h_{c,{\rm lim}}\simeq 10^{-15}$. Multiplying \eqref{eq:n_bound_new} by $m_{\rm PBH}$ gives an upper limit on the local PBH mass density
\begin{equation} 
\rho_{\rm PBH} \equiv m_{\rm PBH} n_{\rm PBH}
\lesssim
\frac{1}{\pi b^{2} v_{\rm PBH} T}
\left(
\frac{h_{c,{\rm lim}}\, b\, v_{\rm PBH}\, c}{2G}
\right)^{2}
\frac{1}{m_{\rm PBH}}
\label{eq:rho_bound_new}
\end{equation}
which scales inversely with PBH mass $\rho_{\rm PBH}^{\rm max}\propto m_{\rm PBH}^{-1}$ and we can then evaluate \eqref{eq:rho_bound_new} for the benchmark parameters as before with a 15 year observing baseline and $h_{c,lim}=10^{-15}$ we obtain
\begin{equation}
\rho_{\rm PBH}^{\rm max}
\simeq 7\times10^{-21}\ {\rm g\,cm^{-3}}
\label{eq:rho_final_result}
\end{equation}
This upper limit remains several orders of magnitude above the canonical local dark matter
density
\begin{equation}
\rho_{\rm DM,local}\simeq 5\times10^{-25}\ {\rm g\,cm^{-3}}
\end{equation}
indicating that current PTA sensitivity to dipolar barycentric kicks does not yet reach the regime required to constrain PBHs at the level of the local dark matter budget.  
Nevertheless, the higher characteristic velocity $v_{\rm PBH}$ considered here slightly strengthens the bound relative to slower encounter scenarios, since the encounter rate $\Gamma\propto v_{\rm PBH}$ increases while the induced kick amplitude retains its $h\propto \Delta v/c$ scaling.  
PTA baselines extending over multiple decades, together with improvements in timing noise and cross correlation analysis, can therefore continue to tighten these constraints.
\\
\\
The parametric dependence in \eqref{eq:rho_bound_new} is interesting because $\rho_{\rm PBH}^{\rm max}\propto m_{\rm PBH}^{-1}$, lowering the assumed PBH mass weakens the bound as for $m_{\rm PBH}=10^{22}~{\rm g}$, one obtains $\rho_{\rm PBH}^{\rm max}\sim10^{-20}~{\rm g~cm^{-3}}$ while if go on increasing is to $m_{\rm PBH}=10^{24}~{\rm g}$ then it tightens to $\rho_{\rm PBH}^{\rm max}\sim10^{-22}~{\rm g~cm^{-3}}$, but this is still well above $\rho_{\rm DM,local}$. Furthermore, the dependence on impact parameter largely cancels out in the random-walk regime because smaller $b$ enhances the kick but simultaneously reduces the geometric encounter rate.
\\
\\
A more physical formulation take a slightly different approach which we shall now discuss. Let us now assume that PBHs make up all of the local dark matter, which means we are fixing
\begin{equation}
\rho_{\rm PBH}=\rho_{\rm DM,local}\simeq5\times10^{-25}~{\rm g~cm^{-3}}
\label{eq:rhoDM}
\end{equation}
The corresponding PBH number density and encounter rate follows as
\begin{equation}
n_{\rm PBH}=\frac{\rho_{\rm PBH}}{m_{\rm PBH}}, \qquad
\Gamma=n_{\rm PBH}\pi b^2 v_{\rm PBH}
\end{equation}
The accumulated barycentric velocity dispersion after time $T$ is then given as
\begin{equation}
\delta v_{\rm PBH,rms}(T)\simeq\sqrt{N(T)}\frac{2Gm_{\rm PBH}}{b v_{\rm PBH}}
=\left[\left(\frac{5\times10^{-25}}{m_{\rm PBH}}\right)\pi b^2 v_{\rm PBH} T\right]^{1/2}\frac{2Gm_{\rm PBH}}{b v_{\rm PBH}}
\end{equation}
this simplifies to
\begin{equation}
\delta v_{\rm PBH,rms}(T)\simeq\left(5\times10^{-25}\pi T\right)^{1/2}\frac{2G}{v_{\rm PBH}^{1/2}}m_{\rm PBH}^{1/2}
\label{eq:dvrms_simplified}
\end{equation}
We see that \eqref{eq:dvrms_simplified} shows that the velocity drift grows with $m_{\rm PBH}^{1/2}$ and is independent of the impact parameter and we can then convert this to a strain-like amplitude to give us 
\begin{equation}
h_{c,{\rm PBH}}\sim\frac{\delta v_{\rm PBH,rms}(T)}{c}
\simeq\left(5\times10^{-25}\pi T\right)^{1/2}\frac{2G}{c v_{\rm PBH}^{1/2}}m_{\rm PBH}^{1/2}
\label{eq:hcPBH_general}
\end{equation}
For $m_{\rm PBH}=10^{23}~{\rm g}$, $v_{\rm PBH}=2\times10^{7}~{\rm cm~s^{-1}}$ and $T=15~{\rm yr}$, this gives
\begin{equation}
h_{c,{\rm PBH}}\simeq3\times10^{-17}
\label{eq:hcPBH_1e23}
\end{equation}
This can be compared with the conservative PTA dipolar sensitivity floor $h_{c,{\rm lim}}\sim10^{-15}$, which means that
\begin{equation}
h_{c,{\rm PBH}}\sim3\times10^{-17}\ll10^{-15}
\end{equation}
meaning that even if PBHs with $m_{\rm PBH}\sim10^{23}~{\rm g}$ make up all of the local dark matter, the induced correlated dipolar signal would still lie below current PTA detectability thresholds.
\\
\\
Since $h_{c,{\rm PBH}}\propto m_{\rm PBH}^{1/2}$, the critical mass where the predicted amplitude reaches current PTA sensitivity is
\begin{equation}
m_{\rm PBH,crit}\sim\left(\frac{10^{-15}}{3\times10^{-17}}\right)^2\times10^{23}~{\rm g}
\simeq10^{26}~{\rm g}
\end{equation}
This corresponds to sub-planetary masses, implying that PBHs heavier than $\sim10^{26}~{\rm g}$ cannot make up the entire dark matter density without producing a detectable PTA dipolar signature and conversely, this means that PBHs lighter than this threshold remain viable as dark matter candidates under present PTA limits. While current sensitivities are not yet competitive with global dark matter bounds, the method has the advantage of probing local, time-resolved gravitational signatures and with ongoing improvements in PTA precision with baselines spanning multiple decades, this approach could evolve into a powerful complementary tool for testing PBHs as dark matter in the Solar neighborhood.
\\
\\
\section{ADAF Flares and Planetary-Mass PBHs}
Advection dominated accretion flows provide a natural pathway for short lived optical flares when a compact object moves through a cold dilute medium and briefly attains an enhanced mass supply \cite{adaf1Narayan:1994xi,adaf2Narayan:1994is,adaf3Narayan:1998ft,adaf4Ali-Haimoud:2016mbv}. In this picture a planetary mass primordial black hole sweeping through the outer Solar System would compress and heat the surrounding gas and dust into a hot optically thin flow in which most of the dissipated energy is advected rather than radiated. The transient luminosity produced in such an ADAF state can reach single visit depths of wide field surveys such as LSST if the instantaneous accretion rate and radiative efficiency are sufficiently large and if the source lies within a few thousand astronomical units \cite{adaf5Ricotti:2007au,adaf10Siraj:2020upy,adaf6Poulin:2017bwe,adaf7Bosch-Ramon:2020pcz,adaf8Martinez:2025git,adaf9Gaggero:2016dpq}. Thus wide fast optical surveys could offer a complementary time domain channel for detecting or constraining nearby planetary mass PBHs
\\
\\
For a PBH of mass $M$ moving with speed $v$ through a medium of mass density $\rho$ and sound speed $c_s$ the Bondi accretion rate is
\begin{equation}
\dot M_{\rm BHL}=\frac{4\pi G^2 M^2 \rho}{\left(v^2+c_s^2\right)^{3/2}}
\end{equation}
This is complimented with a fiducial accretion luminosity is $L=\epsilon\,\dot M_{\rm BHL} c^2$ with radiative efficiency $\epsilon$ appropriate for a hot ADAF. At an observer distance $D$ the bolometric energy flux is
\begin{equation}
F=\frac{L}{4\pi D^2}=\frac{\epsilon\,\dot M_{\rm BHL} c^2}{4\pi D^2}
\end{equation}
A passing icy body of bulk density $\rho_{\rm obj}$ and radius $s$ can be tidally processed if it approaches within the tidal radius
\begin{equation}
R_{\rm TD}\simeq \left(\frac{M}{\rho_{\rm obj}}\right)^{1/3}
\end{equation}
This means that an effective geometric cross section for disruptive encounters is $\sigma_{\rm TD}\sim \pi R_{\rm TD}^2$, up to gravitational focusing corrections near the PBH. The local number density of outer solar system objects larger than $s$ is $n_{\rm OC}(>s)$ the PBH object encounter rate is given as
\begin{equation}
\Gamma_{\rm coll}\sim n_{\rm OC}(>s)\,\sigma_{\rm TD}\,v
\end{equation}
During each encounter the sublimation and shredding of the body can drive a transient $\dot M$ onto the PBH and so a practical detectability condition for a single flare with duration $\tau$ which is set by the residence time near $R_{\rm TD}$ is given as
\begin{equation}
F \gtrsim F_{\rm lim}
\end{equation}
where $F_{\rm lim}$ refers to a single visit optical threshold and also equivalently a convenient detectability radius is given as
\begin{equation}
D_{\rm det}\simeq \left(\frac{\epsilon\,\dot M_{\rm BHL} c^2}{4\pi F_{\rm lim}}\right)^{1/2}
\end{equation}
Using $M = 3\times10^{28} g$, $v_{\rm PBH} = 2\times10^{7}$, $\rho = 10^{-24}\ {\rm g\,cm^{-3}}$ and $\epsilon =10^{-2}-10^{-3}$, we have the Bondi–Hoyle–Lyttleton accretion rate for an encounter as
\begin{equation}
\dot{M}_{\rm BHL}
 = \frac{4\pi G^{2} M^{2} \rho}{(v_{\rm PBH}^{2}+c_{s}^{2})^{3/2}} 
 \approx 6.3\times10^{-3}\ {\rm g\,s^{-1}}
\end{equation}
This gives an ADAF luminosity
\begin{equation}
L = \epsilon\,\dot{M}_{\rm BHL} c^{2} 
\approx (5.7\times10^{15}\text{-}5.7\times10^{16})\ {\rm erg\,s^{-1}}
\end{equation}
For an outer solar system icy body of bulk density $\rho_{\rm obj}=0.5\ {\rm g\,cm^{-3}}$, the tidal disruption radius is
\begin{equation}
R_{\rm TD}\simeq \left(\frac{M}{\rho_{\rm obj}}\right)^{1/3}
\approx 4\times10^{9}\ {\rm cm}
\end{equation}
leading to an effective tidal cross section
\begin{equation}
\sigma_{\rm TD}\sim \pi R_{\rm TD}^{2}
\approx 5\times10^{19}\ {\rm cm^{2}}
\end{equation}
If we now adopt a conservative single–visit optical threshold of 
$F_{\rm lim}\sim 3\times10^{-15}\ {\rm erg\,cm^{-2}\,s^{-1}}$ then
the corresponding detectability distance of a flare is
\begin{equation}
D_{\rm det}
\simeq \left(\frac{L}{4\pi F_{\rm lim}}\right)^{1/2}
\approx (3.9\times10^{14}\text{-}1.2\times10^{15})\ {\rm cm}
\approx (26\text{-}82)\ {\rm au}
\end{equation}
Thus for PBHs moving at $v_{\rm PBH}\sim200~{\rm km\,s^{-1}}$, ADAF powered flares naturally select the Kuiper Belt and scattered disk region as the most promising search volume. In this regime the accretion flow becomes short–lived and sharply peaked in luminosity, which creates a clean, localized transient signature that wide–field surveys can effectively isolate against background variability. This sharpening of the emission zone turns the inner $(20\text{--}100)$ au of the Solar System into a well-defined laboratory where even a single interaction would stand out prominently in deep optical data. PBHs located farther out in the Oort Cloud would produce fainter events, but the inner-system channel becomes a powerful and well controlled probe for fast, planetary–mass PBHs.
\\
\\
We now consider the event rate for this channel where instead of considering just a per-PBH collision rate, we calculate the total number of PBH-object encounters per unit time in the shell which can follow directly from kinetic theory. The reaction rate density is  
$n_{\rm PBH}\,n_{\rm OC}\,\sigma_{\rm TD}\,v_{\rm PBH}$, so integrating over the shell volume $V$ gives
\begin{equation}
\Gamma_{\rm all} \;=\; \int_V n_{\rm PBH}\,n_{\rm OC}\,\sigma_{\rm TD}\,v_{\rm PBH}\,{\rm d}V 
\;=\; n_{\rm PBH}\,n_{\rm OC}\,\sigma_{\rm TD}\,v_{\rm PBH}\,V 
\;=\; n_{\rm PBH}\,N_{\rm OC}(>s)\,\sigma_{\rm TD}\,v_{\rm PBH}
\end{equation}
where $n_{\rm OC}=N_{\rm OC}(>s)/V$ and the volume cancels out and then, using $n_{\rm PBH}=f_{\rm PBH}\rho_{\rm DM}/M$ and $\sigma_{\rm TD}\simeq \pi R_{\rm TD}^2$ with 
$R_{\rm TD}\simeq (M/\rho_{\rm obj})^{1/3}$, we obtain the compact form
\begin{equation}
\Gamma_{\rm all}(>s) \;=\; f_{\rm PBH}\,\frac{\rho_{\rm DM}}{M}\,N_{\rm OC}(>s)\,\pi \left(\frac{M}{\rho_{\rm obj}}\right)^{\!2/3} v_{\rm PBH}
\;=\; f_{\rm PBH}\,N_{\rm OC}(>s)\,\pi\,v_{\rm PBH}\,\rho_{\rm DM}\,\rho_{\rm obj}^{-2/3}\,M^{-1/3}
\label{eq:GammaAllCompact}
\end{equation}
This makes the scalings explicit as $\Gamma_{\rm all}\propto f_{\rm PBH}\,N_{\rm OC}\,v_{\rm PBH}\,M^{-1/3}$ for fixed $\rho_{\rm DM}$ and $\rho_{\rm obj}$. Now we adop adopting similar values as before for the PBHs with $\rho_{\rm obj}=0.5\ {\rm g\,cm^{-3}} , N_{\rm OC}(>1{\rm\,km})\simeq 10^{11}$, we have \begin{equation}
    R_{\rm TD}\simeq (M/\rho_{\rm obj})^{1/3}\approx 5\times10^{19}\ {\rm cm^2}
\end{equation} Inserting these into \eqref{eq:GammaAllCompact} gives us
\begin{equation}
\Gamma_{\rm all}(>1{\rm\,km}) \approx\; 5.3\times10^{-8}\,f_{\rm PBH}\ {\rm yr^{-1}}
\end{equation}
Even if PBHs make up all of the local dark matter ($f_{\rm PBH}=1$), this corresponds to one disruptive PBH-Oort object encounter in the entire shell roughly every 
$\sim 2\times 10^{7}$ years. The linear $v_{\rm PBH}$ dependence in \eqref{eq:GammaAllCompact} slightly increases $\Gamma_{\rm all}$ at higher speeds, but this does not offset the strong reduction in ADAF luminosity (relevant for detectability) which scales as $\dot M_{\rm BHL}\propto v_{\rm PBH}^{-3}$. We see that while intrinsic geometric collision rate from the whole PBH population is fixed by the local dark-matter density and scales as $M^{-1/3}$, the observable flare yield remains dominated by rare, nearby encounters in the inner $(20$-$100)$ au where flux limits can be met.
\\
\\
Because ADAF powered flares scale with the instantaneous local supply of gas and tidally liberated debris and because detectability scales as $D_{\rm det}\propto (\epsilon\,\dot M_{\rm BHL})^{1/2}$, we see that nearby PBHs with masses being a few solar masses $M_0$ can in principle be tested with LSST class surveys out to $\sim 10^{2}$ au while more distant objects fall below single visit depth. This makes the outer Solar System quite a viable arena to constrain the existence of planetary mass PBHs where many cosmological and galactic probes are either insensitive or highly model dependent as discussed before. The flare channel is hence very unique in targeting heavy PBHs that are otherwise difficult to test locally providing a time domain complement to the low frequency barycentric perturbation methods developed in this work and it also motivates systematic searches for fast optical transients consistent with ADAF like spectra and durations in broader wide field survey data.
\\
\\
\section{Conclusions}
In this work we have proposed and explored a new class of Solar System scale probes for PBHs across a wide range of masses, giving two complementary approaches. For PBHs having masses in the asteroid to dwarf planet range, we examined how PTAs can be used to detect or constrain the correlated dipolar timing signatures that arise from cumulative gravitational kicks imparted to the Solar System barycenter. For planetary mass PBHs, we developed a framework to estimate the luminosity, detectability and event rate of accretion flares arising from ADAF triggered by the tidal disruption and sublimation of icy bodies in the outer Solar System. Together, these analyses provide us a unified picture of how PTA timing and wide-field optical surveys can serve as quite natural local, time-domain observatories for testing PBHs as DM candidates.
\\
\\
These Solar System probes are novel as unlike more traditional PBH searches that rely on cosmological or galactic-scale effects, our approach focuses on local, time-resolved gravitational and accretion signatures directly measurable within the Solar neighborhood. The PTA based method connects low frequency timing residuals to small but coherent accelerations of the Solar System barycenter and thus offering sensitivity to asteroid-mass PBHs. The ADAF flare mechanism on the other hand, targets PBHs in the planetary mass range, which are often beyond the reach of other detection procedures. These methods jointly span mass ranges that are currently the least constrained by cosmological observations, providing an independent and complementary window into PBH phenomenology.
\\
\\
The results of this work can also open up promising pathways for future explorations, which can include extending PTA analysis to include long term dipolar correlations combined with improved time baselines. This will result in enhancing the sensitivity to the diffusive barycentric motion expected from PBHs making up the local dark matter density. At the same time, systematic searches for fast optical transients consistent with ADAF like spectral signatures could empirically constrain or perhaps even discover nearby planetary mass PBHs. In this way, Solar System physics itself becomes an active laboratory for exploring the nature of dark matter and for searching for the elusive relics of the early Universe.
\\
\\
\section*{Acknowledgements}
OT was supported in part by the Vanderbilt Discovery Alliance Fellowship. AL was supported in part by the Black Hole Initiative, which is funded by GBMF and JTF. OT especially thanks the hospitality of Harvard's Center for Astrophysics, as the ideas for this work originated during their stay there.

\bibliography{references}

\begin{thebibliography}{10}

\bibitem{dm11rubin1970rotation}
Vera~C Rubin and W~Kent Ford~Jr.
\newblock Rotation of the andromeda nebula from a spectroscopic survey of emission regions.
\newblock {\em Astrophysical Journal, vol. 159, p. 379}, 159:379, 1970.

\bibitem{dm1Cirelli:2024ssz}
Marco Cirelli, Alessandro Strumia, and Jure Zupan.
\newblock {Dark Matter}.
\newblock 6 2024.

\bibitem{dm2Arbey:2021gdg}
A.~Arbey and F.~Mahmoudi.
\newblock {Dark matter and the early Universe: a review}.
\newblock {\em Prog. Part. Nucl. Phys.}, 119:103865, 2021.

\bibitem{dm3Balazs:2024uyj}
Csaba Balazs, Torsten Bringmann, Felix Kahlhoefer, and Martin White.
\newblock {A Primer on Dark Matter}.
\newblock 11 2024.

\bibitem{dm4Eberhardt:2025caq}
Andrew Eberhardt and Elisa G.~M. Ferreira.
\newblock {Ultralight fuzzy dark matter review}.
\newblock 7 2025.

\bibitem{dm5Bozorgnia:2024pwk}
Nassim Bozorgnia, Joseph Bramante, James~M. Cline, David Curtin, David McKeen, David~E. Morrissey, Adam Ritz, Simon Viel, Aaron~C. Vincent, and Yue Zhang.
\newblock {Dark Matter Candidates and Searches}.
\newblock 9 2024.

\bibitem{dm6Misiaszek:2023sxe}
Marcin Misiaszek and Nicola Rossi.
\newblock {Direct Detection of Dark Matter: A Critical Review}.
\newblock {\em Symmetry}, 16(2):201, 2024.

\bibitem{dm7OHare:2024nmr}
Ciaran A.~J. O'Hare.
\newblock {Cosmology of axion dark matter}.
\newblock {\em PoS}, COSMICWISPers:040, 2024.

\bibitem{dm8Adhikari:2022sbh}
Susmita Adhikari et~al.
\newblock {Astrophysical Tests of Dark Matter Self-Interactions}.
\newblock 7 2022.

\bibitem{dm9Miller:2025yyx}
Andrew~L. Miller.
\newblock {Gravitational wave probes of particle dark matter: a review}.
\newblock 3 2025.

\bibitem{pbh1pbhzel1966hypothesis}
Ya~B Zel'Dovich and ID~Novikov.
\newblock The hypothesis of cores retarded during expansion and the hot cosmological model.
\newblock {\em Astronomicheskii Zhurnal}, 43:758, 1966.

\bibitem{pbh2hawking1971gravitationally}
Stephen Hawking.
\newblock Gravitationally collapsed objects of very low mass.
\newblock {\em Monthly Notices of the Royal Astronomical Society}, 152(1):75--78, 1971.

\bibitem{pbh3carr1974black}
Bernard~J Carr and Stephen~W Hawking.
\newblock Black holes in the early universe.
\newblock {\em Monthly Notices of the Royal Astronomical Society}, 168(2):399--415, 1974.

\bibitem{pbh4carr1975primordial}
Bernard~J Carr.
\newblock The primordial black hole mass spectrum, 1975.

\bibitem{pbh5chapline1975cosmological}
George~F Chapline.
\newblock Cosmological effects of primordial black holes.
\newblock {\em Nature}, 253(5489):251--252, 1975.

\bibitem{pbh6hawking1975particle}
Stephen~W Hawking.
\newblock Particle creation by black holes.
\newblock {\em Communications in mathematical physics}, 43(3):199--220, 1975.

\bibitem{pbh7hawking1974black}
Stephen~W Hawking.
\newblock Black hole explosions?
\newblock {\em Nature}, 248(5443):30--31, 1974.

\bibitem{pbh8khlopov1980primordial}
M~Yu Khlopov and AG~Polnarev.
\newblock Primordial black holes as a cosmological test of grand unification.
\newblock {\em Physics Letters. Section B: Nuclear, Elementary Particle and High-Energy Physics}, 97(3-4):383--387, 1980.

\bibitem{pbh9polnarev1985cosmology}
Alexandre~G Polnarev and M~Yu Khlopov.
\newblock Cosmology, primordial black holes, and supermassive particles.
\newblock {\em Soviet Physics Uspekhi}, 28(3):213, 1985.

\bibitem{pbh10khlopov2010primordial}
Maxim~Yu Khlopov.
\newblock Primordial black holes.
\newblock {\em Research in Astronomy and Astrophysics}, 10(6):495, 2010.

\bibitem{pbh11carr2016primordial}
Bernard Carr, Florian K{\"u}hnel, and Marit Sandstad.
\newblock Primordial black holes as dark matter.
\newblock {\em Physical Review D}, 94(8):083504, 2016.

\bibitem{dm10Trivedi:2025vry}
Oem Trivedi and Abraham Loeb.
\newblock {Could planck star remnants be dark matter?}
\newblock {\em Phys. Dark Univ.}, 49:102003, 2025.

\bibitem{pbh12carr2020primordial}
Bernard Carr and Florian K{\"u}hnel.
\newblock Primordial black holes as dark matter: recent developments.
\newblock {\em Annual Review of Nuclear and Particle Science}, 70(1):355--394, 2020.

\bibitem{pbh13carr2024observational}
BJ~Carr, Sebastien Clesse, Juan Garcia-Bellido, MRS Hawkins, and Florian K{\"u}hnel.
\newblock Observational evidence for primordial black holes: A positivist perspective.
\newblock {\em Physics Reports}, 1054:1--68, 2024.

\bibitem{pbh14green2024primordial}
Anne~M Green.
\newblock Primordial black holes as a dark matter candidate-a brief overview.
\newblock {\em Nuclear Physics B}, 1003:116494, 2024.

\bibitem{test1Cai:2022kbp}
Rong-Gen Cai, Tan Chen, Shao-Jiang Wang, and Xing-Yu Yang.
\newblock {Gravitational microlensing by dressed primordial black holes}.
\newblock {\em JCAP}, 03:043, 2023.

\bibitem{test2Niikura:2019kqi}
Hiroko Niikura, Masahiro Takada, Shuichiro Yokoyama, Takahiro Sumi, and Shogo Masaki.
\newblock {Constraints on Earth-mass primordial black holes from OGLE 5-year microlensing events}.
\newblock {\em Phys. Rev. D}, 99(8):083503, 2019.

\bibitem{test3Tashiro:2008sf}
Hiroyuki Tashiro and Naoshi Sugiyama.
\newblock {Constraints on Primordial Black Holes by Distortions of Cosmic Microwave Background}.
\newblock {\em Phys. Rev. D}, 78:023004, 2008.

\bibitem{test4Graham:2023unf}
Peter~W. Graham and Harikrishnan Ramani.
\newblock {Constraints on dark matter from dynamical heating of stars in ultrafaint dwarfs. I. MACHOs and primordial black holes}.
\newblock {\em Phys. Rev. D}, 110(7):075011, 2024.

\bibitem{test5Auffinger:2022khh}
J{\'e}r{\'e}my Auffinger.
\newblock {Primordial black hole constraints with Hawking radiation{\textemdash}A review}.
\newblock {\em Prog. Part. Nucl. Phys.}, 131:104040, 2023.

\bibitem{test6Klipfel:2025bvh}
Alexandra~P. Klipfel, Peter Fisher, and David~I. Kaiser.
\newblock {Hawking Radiation Signatures from Primordial Black Holes Transiting the Inner Solar System: Prospects for Detection}.
\newblock 6 2025.

\bibitem{test7Niikura:2017zjd}
Hiroko Niikura et~al.
\newblock {Microlensing constraints on primordial black holes with Subaru/HSC Andromeda observations}.
\newblock {\em Nature Astron.}, 3(6):524--534, 2019.

\bibitem{test8wyrzykowski2011ogle}
{\L}~Wyrzykowski, S~Koz{\l}owski, J~Skowron, A~Udalski, MK~Szyma{\'n}ski, M~Kubiak, G~Pietrzy{\'n}ski, I~Soszy{\'n}ski, O~Szewczyk, K~Ulaczyk, et~al.
\newblock The ogle view of microlensing towards the magellanic clouds--iii. ruling out subsolar machos with the ogle-iii lmc data.
\newblock {\em Monthly Notices of the Royal Astronomical Society}, 413(1):493--508, 2011.

\bibitem{test9EROS:2002flm}
C.~Afonso et~al.
\newblock {Limits on galactic dark matter with 5 years of EROS SMC data}.
\newblock {\em Astron. Astrophys.}, 400:951--956, 2003.

\bibitem{haw1Carr:2009jm}
B.~J. Carr, Kazunori Kohri, Yuuiti Sendouda, and Jun'ichi Yokoyama.
\newblock {New cosmological constraints on primordial black holes}.
\newblock {\em Phys. Rev. D}, 81:104019, 2010.

\bibitem{haw2Boudaud:2018hqb}
Mathieu Boudaud and Marco Cirelli.
\newblock {Voyager 1 $e^\pm$ Further Constrain Primordial Black Holes as Dark Matter}.
\newblock {\em Phys. Rev. Lett.}, 122(4):041104, 2019.

\bibitem{haw3Laha:2019ssq}
Ranjan Laha.
\newblock {Primordial Black Holes as a Dark Matter Candidate Are Severely Constrained by the Galactic Center 511 keV $\gamma$ -Ray Line}.
\newblock {\em Phys. Rev. Lett.}, 123(25):251101, 2019.

\bibitem{acc1Agius:2024ecw}
Dominic Agius, Rouven Essig, Daniele Gaggero, Francesca Scarcella, Gregory Suczewski, and Mauro Valli.
\newblock {Feedback in the dark: a critical examination of CMB bounds on primordial black holes}.
\newblock {\em JCAP}, 07:003, 2024.

\bibitem{acc2Serpico:2020ehh}
Pasquale~D. Serpico, Vivian Poulin, Derek Inman, and Kazunori Kohri.
\newblock {Cosmic microwave background bounds on primordial black holes including dark matter halo accretion}.
\newblock {\em Phys. Rev. Res.}, 2(2):023204, 2020.

\bibitem{acc3Piga:2022ysp}
Lorenzo Piga, Matteo Lucca, Nicola Bellomo, Valent{\`\i} Bosch-Ramon, Sabino Matarrese, Alvise Raccanelli, and Licia Verde.
\newblock {The effect of outflows on CMB bounds from Primordial Black Hole accretion}.
\newblock {\em JCAP}, 12:016, 2022.

\bibitem{pta1NANOGrav:2023gor}
Gabriella Agazie et~al.
\newblock {The NANOGrav 15 yr Data Set: Evidence for a Gravitational-wave Background}.
\newblock {\em Astrophys. J. Lett.}, 951(1):L8, 2023.

\bibitem{pta2NANOGrav:2023hde}
Gabriella Agazie et~al.
\newblock {The NANOGrav 15 yr Data Set: Observations and Timing of 68 Millisecond Pulsars}.
\newblock {\em Astrophys. J. Lett.}, 951(1):L9, 2023.

\bibitem{pta3Xu:2023wog}
Heng Xu et~al.
\newblock {Searching for the Nano-Hertz Stochastic Gravitational Wave Background with the Chinese Pulsar Timing Array Data Release I}.
\newblock {\em Res. Astron. Astrophys.}, 23(7):075024, 2023.

\bibitem{pta4Antoniadis:2022pcn}
J.~Antoniadis et~al.
\newblock {The International Pulsar Timing Array second data release: Search for an isotropic gravitational wave background}.
\newblock {\em Mon. Not. Roy. Astron. Soc.}, 510(4):4873--4887, 2022.

\bibitem{pta5Zic:2023gta}
Andrew Zic et~al.
\newblock {The Parkes Pulsar Timing Array third data release}.
\newblock {\em Publ. Astron. Soc. Austral.}, 40:e049, 2023.

\bibitem{pta6Reardon:2023gzh}
Daniel~J. Reardon et~al.
\newblock {Search for an Isotropic Gravitational-wave Background with the Parkes Pulsar Timing Array}.
\newblock {\em Astrophys. J. Lett.}, 951(1):L6, 2023.

\bibitem{pta7EPTA:2023sfo}
J.~Antoniadis et~al.
\newblock {The second data release from the European Pulsar Timing Array - I. The dataset and timing analysis}.
\newblock {\em Astron. Astrophys.}, 678:A48, 2023.

\bibitem{Loeb:2024ekw}
Abraham Loeb.
\newblock {Pulsar Timing Noise from Brownian Motion of the Sun}.
\newblock {\em Astrophys. J. Lett.}, 968(2):L27, 2024.

\bibitem{adaf1Narayan:1994xi}
Ramesh Narayan and In-su Yi.
\newblock {Advection dominated accretion: A Selfsimilar solution}.
\newblock {\em Astrophys. J. Lett.}, 428:L13, 1994.

\bibitem{adaf2Narayan:1994is}
Ramesh Narayan and Insu Yi.
\newblock {Advection dominated accretion: Underfed black holes and neutron stars}.
\newblock {\em Astrophys. J.}, 452:710, 1995.

\bibitem{adaf3Narayan:1998ft}
Ramesh Narayan, Rohan Mahadevan, and Eliot Quataert.
\newblock {Advection - dominated accretion around black holes}.
\newblock 3 1998.

\bibitem{adaf4Ali-Haimoud:2016mbv}
Yacine Ali-Ha{\"\i}moud and Marc Kamionkowski.
\newblock {Cosmic microwave background limits on accreting primordial black holes}.
\newblock {\em Phys. Rev. D}, 95(4):043534, 2017.

\bibitem{adaf5Ricotti:2007au}
Massimo Ricotti, Jeremiah~P. Ostriker, and Katherine~J. Mack.
\newblock {Effect of Primordial Black Holes on the Cosmic Microwave Background and Cosmological Parameter Estimates}.
\newblock {\em Astrophys. J.}, 680:829, 2008.

\bibitem{adaf10Siraj:2020upy}
Amir Siraj and Abraham Loeb.
\newblock {Searching for Black Holes in the Outer Solar System with LSST}.
\newblock {\em Astrophys. J. Lett.}, 898(1):L4, 2020.

\bibitem{adaf6Poulin:2017bwe}
Vivian Poulin, Pasquale~D. Serpico, Francesca Calore, Sebastien Clesse, and Kazunori Kohri.
\newblock {CMB bounds on disk-accreting massive primordial black holes}.
\newblock {\em Phys. Rev. D}, 96(8):083524, 2017.

\bibitem{adaf7Bosch-Ramon:2020pcz}
Valenti Bosch-Ramon and Nicola Bellomo.
\newblock {Mechanical feedback effects on primordial black hole accretion}.
\newblock {\em Astron. Astrophys.}, 638:A132, 2020.

\bibitem{adaf8Martinez:2025git}
Javier~Rodrigo Martinez, Valenti Bosch-Ramon, Florencia~Laura Vieyro, and Santiago del Palacio.
\newblock {Probing the detectability of electromagnetic signatures from Galactic isolated black holes}.
\newblock {\em Astron. Astrophys.}, 700:A49, 2025.

\bibitem{adaf9Gaggero:2016dpq}
Daniele Gaggero, Gianfranco Bertone, Francesca Calore, Riley M.~T. Connors, Mark Lovell, Sera Markoff, and Emma Storm.
\newblock {Searching for Primordial Black Holes in the radio and X-ray sky}.
\newblock {\em Phys. Rev. Lett.}, 118(24):241101, 2017.

\end{thebibliography}
\bibliographystyle{unsrt}

\end{document}